\documentclass[12pt]{article}
\usepackage{lmodern}
\usepackage{lmodern}
\usepackage[T1]{fontenc}
\usepackage[utf8]{inputenc}
\usepackage{color}
\usepackage{amsmath}
\usepackage{amssymb}
\usepackage{graphicx}
\usepackage[authoryear]{natbib}
\usepackage[bookmarks=false,
 breaklinks=false,pdfborder={0 0 1},colorlinks=true]
 {hyperref}
\hypersetup{pdftitle={Integrating RCTs, RWD, and AI/ML: A Statistical Perspective on Next-Generation Evidence Synthesis},
 pdfauthor={Shu Yang; Margaret Gamalo; Haoda Fu},
 pdfkeywords={real-world evidence; causal inference; machine learning; clinical trials; evidence synthesis; regulatory science},
 linkcolor={blue},filecolor={Maroon},citecolor={Blue},urlcolor={Blue},pdfcreator={LaTeX via pandoc}}

\makeatletter
\@ifundefined{date}{}{\date{}}

\usepackage{setspace}
\AtBeginEnvironment{thebibliography}{\setstretch{0.8}}

\usepackage{iftex}
\ifPDFTeX
  \usepackage{textcomp}
\else 
  \usepackage{unicode-math}
\defaultfontfeatures{Scale=MatchLowercase}
  \defaultfontfeatures[\rmfamily]{Ligatures=TeX,Scale=1}
\fi
\ifPDFTeX\else  
\fi
\IfFileExists{upquote.sty}{\usepackage{upquote}}{}
\IfFileExists{microtype.sty}{
  \usepackage[]{microtype}
  \UseMicrotypeSet[protrusion]{basicmath} 
}{}

\@ifundefined{KOMAClassName}{
  \IfFileExists{parskip.sty}{%
    \usepackage{parskip}
  }{
    \setlength{\parindent}{0pt}
    \setlength{\parskip}{6pt plus 2pt minus 1pt}}
}{
  \KOMAoptions{parskip=half}}

\usepackage[dvipsnames,svgnames,x11names]{xcolor}
\ifx\paragraph\undefined\else
  \let\oldparagraph\paragraph
  \renewcommand{\paragraph}{
    \@ifstar
      \xxxParagraphStar
      \xxxParagraphNoStar
  }
  \newcommand{\xxxParagraphStar}[1]{\oldparagraph*{#1}\mbox{}}
  \newcommand{\xxxParagraphNoStar}[1]{\oldparagraph{#1}\mbox{}}
\fi
\ifx\subparagraph\undefined\else
  \let\oldsubparagraph\subparagraph
  \renewcommand{\subparagraph}{
    \@ifstar
      \xxxSubParagraphStar
      \xxxSubParagraphNoStar
  }
  \newcommand{\xxxSubParagraphStar}[1]{\oldsubparagraph*{#1}\mbox{}}
  \newcommand{\xxxSubParagraphNoStar}[1]{\oldsubparagraph{#1}\mbox{}}
\fi

\renewcommand{\oldparagraph}{\@startsection{paragraph}{4}{\z@}%
  {\parskip}%
  {-1em}%
  {\normalfont\normalsize\bfseries}}

\usepackage{longtable}
\usepackage{booktabs}\usepackage{array}\usepackage{calc}
\usepackage{etoolbox}

\patchcmd\longtable{\par}{\if@noskipsec\mbox{}\fi\par}{}{}

\IfFileExists{footnotehyper.sty}{\usepackage{footnotehyper}}{\usepackage{footnote}}
\makesavenoteenv{longtable}

\def\maxwidth{\ifdim\Gin@nat@width>\linewidth\linewidth\else\Gin@nat@width\fi}
\def\maxheight{\ifdim\Gin@nat@height>\textheight\textheight\else\Gin@nat@height\fi}

\setkeys{Gin}{width=\maxwidth,height=\maxheight,keepaspectratio}

\def\fps@figure{htbp}

\@ifpackageloaded{caption}{}{\usepackage{caption}}
\AtBeginDocument{%
\ifdefined\contentsname
  \renewcommand*\contentsname{Table of contents}
\else
  \newcommand\contentsname{Table of contents}
\fi
\ifdefined\listfigurename
  \renewcommand*\listfigurename{List of Figures}
\else
  \newcommand\listfigurename{List of Figures}
\fi
\ifdefined\listtablename
  \renewcommand*\listtablename{List of Tables}
\else
  \newcommand\listtablename{List of Tables}
\fi
\ifdefined\figurename
  \renewcommand*\figurename{Figure}
\else
  \newcommand\figurename{Figure}
\fi
\ifdefined\tablename
  \renewcommand*\tablename{Table}
\else
  \newcommand\tablename{Table}
\fi
}
\@ifpackageloaded{float}{}{\usepackage{float}}
\floatstyle{ruled}
\@ifundefined{c@chapter}{\newfloat{codelisting}{h}{lop}}{\newfloat{codelisting}{h}{lop}[chapter]}
\floatname{codelisting}{Listing}

\@ifpackageloaded{caption}{}{\usepackage{caption}}
\@ifpackageloaded{subcaption}{}{\usepackage{subcaption}}

\ifLuaTeX
  \usepackage{selnolig}
\fi

\usepackage{bookmark}

\IfFileExists{xurl.sty}{\usepackage{xurl}}{} 
\newcommand{\anon}{1}

\usepackage{setspace}

\makeatother

\begin{document}
\singlespacing


\if1\anon { 
\title{\textbf{Integrating RCTs, RWD, and AI/ML: }\textbf{A
Statistical Perspective on}\textbf{ Next-Generation Evidence Synthesis}}
\author{Shu Yang\thanks{Department of Statistics, North Carolina State University, North Carolina
27695, U.S.A. Email: syang24@ncsu.edu}, Margaret Gamalo\thanks{VP and Statistics Head, Inflammation, Immunology \& Specialty Care,
Pfizer}, Haoda Fu\thanks{Head of Exploratory Biostatistics, Amgen}}

\maketitle
} \fi

\if0\anon { \bigskip{}
 \bigskip{}
 \bigskip{}

\begin{center}
{\LARGE\textbf{Integrating RCTs, RWD, and AI/ML: }}{\LARGE\textbf{A
Statistical Perspective on}}{\LARGE\textbf{ Next-Generation Evidence
Synthesis}}\textbf{ } 
\par\end{center}

\medskip{}
 } \fi

\bigskip{}

\begin{abstract}
Randomized controlled trials (RCTs) have been the cornerstone of clinical
evidence; however, their cost, duration, and restrictive eligibility
criteria limit power and external validity. Studies using real-world
data (RWD), historically considered less reliable for establishing
causality, are now recognized as an important source of real-world
evidence (RWE). In parallel, artificial intelligence and machine learning
(AI/ML) are increasingly used throughout the drug development
process, providing scalability and flexibility but also presenting
challenges in interpretability and statistical rigor. This Perspective
argues that the future of evidence generation will not depend on RCTs
versus RWD, or statistics versus AI/ML, but on their
principled integration under \textit{a statistical
evidence framework} that clarifies estimands, evaluates
data fitness, controls bias, quantifies uncertainty, and determines
when evidence is strong enough to support decisions. Building on the Causal Roadmap for high-quality real-world evidence, we present a six-step
statistical roadmap for integrative evidence synthesis and organize
the discussion around five core questions that statisticians must
confront: when RWD is fit for causal use and when it is not; what
AI genuinely contributes across the evidence lifecycle; what remains
distinctly statistical and indispensable; how to build trustworthy
end-to-end evidence systems in pharmaceutical, regulatory, and industry
settings; and how statistical training should evolve. We highlight
key applications of integrative evidence synthesis, including AI-assisted
recruitment and prognostic covariate adjustment within RCTs, target
trial emulation using observational data, synthetic controls and generative
AI, hybrid controlled trials with statistical borrowing safeguards,
extending short-term RCTs with long-term real-world follow-up, and
transporting trial findings to broader target populations. By uniting
statistical rigor with AI/ML innovation, integrative approaches can
produce robust, transparent, and policy-relevant evidence, making
them a key component of modern regulatory science. 
\end{abstract}
 \vfill{}

\newpage\doublespacing


\section{RWD/E and AI/ML in Clinical and Regulatory Studies }

\label{sec:Introduction}

\paragraph*{Rethinking the evidence hierarchy.}
Randomized controlled trials (RCTs) have long been regarded as the
cornerstone of evidence in medicine, providing the highest level of
internal validity by mitigating confounding
and bias. However, RCTs are often expensive and time-consuming,
and rigid eligibility criteria can limit their generalizability. Real-world data (RWD) includes routinely
collected patient health information outside of controlled experimental
settings, such as electronic health records, insurance claims, disease
registries, and data from digital health technologies \citep{FDA2018}.
Historically, RWD was viewed as less reliable and not
fit for purpose because it was not originally intended to serve as
primary evidence, raising concerns about bias, unmeasured confounding,
and data quality. Nonetheless, RWD is now widely recognized as an
important and complementary source of evidence \citep{hampson2018real}.
When designed and analyzed rigorously, RWD studies have the potential
to generate real-world evidence (RWE) capable of addressing questions
beyond the scope of RCTs.

\paragraph*{Policy as a catalyst.}
The widespread adoption of electronic health records
was initiated through the ``Meaningful Use'' program \citep{Blumenthal2010MeaningfulUse},
which laid the foundation for large-scale digital health data collection.
In many respects, the 21st Century Cures Act represented a policy
response to this rapid data expansion, further catalyzing the integration
of real-world data and evidence into U.S.\ FDA regulatory decision-making
\citep{kesselheim2017new}. In 2018, the FDA established an RWE Framework
\citep{FDA2018} to formalize pathways for its use in new indications,
safety surveillance, and primary evidence in specific contexts. This
marked a turning point in recognizing RWE as an integral component
of evidence-based regulatory science.

Evidence generation, whether serving as primary
or supportive evidence, now spans a continuum, from RCTs augmented
by RWD elements to fully observational studies \citep{concato2022real}.
At the design stage, RWD increasingly supports RCTs by informing feasibility
assessments, participant eligibility, and recruitment strategies through
analysis of electronic health records and insurance claims. Pragmatic
trials embed randomization within routine care to enhance generalizability
while preserving internal validity. At the observational
end, registries and external-control designs provide evidence where
RCTs are infeasible, particularly in rare diseases and urgent public
health settings. Collectively, this evolving landscape demonstrates
that RWD/E is no longer peripheral but central to modern evidence
generation.

RWE is already influencing regulatory approvals. Between 2019 and 2021,
RWE contributed to $31\%$ of FDA approvals of new drugs and biologics,
supporting therapeutic context, safety monitoring, and effectiveness
assessments \citep{bachinger2025real}. The global picture is similar:
regulatory guidance from the U.S. FDA \citep{fda2024ehrclaims},
the European Medicines Agency, and the U.K. Medicines and Healthcare products
Regulatory Agency, together with pilot programs in Japan and China, demonstrates
a worldwide movement toward integrating RWD/E into regulatory science.
\begin{figure}[htbp]
\centering
\includegraphics[width=\textwidth]{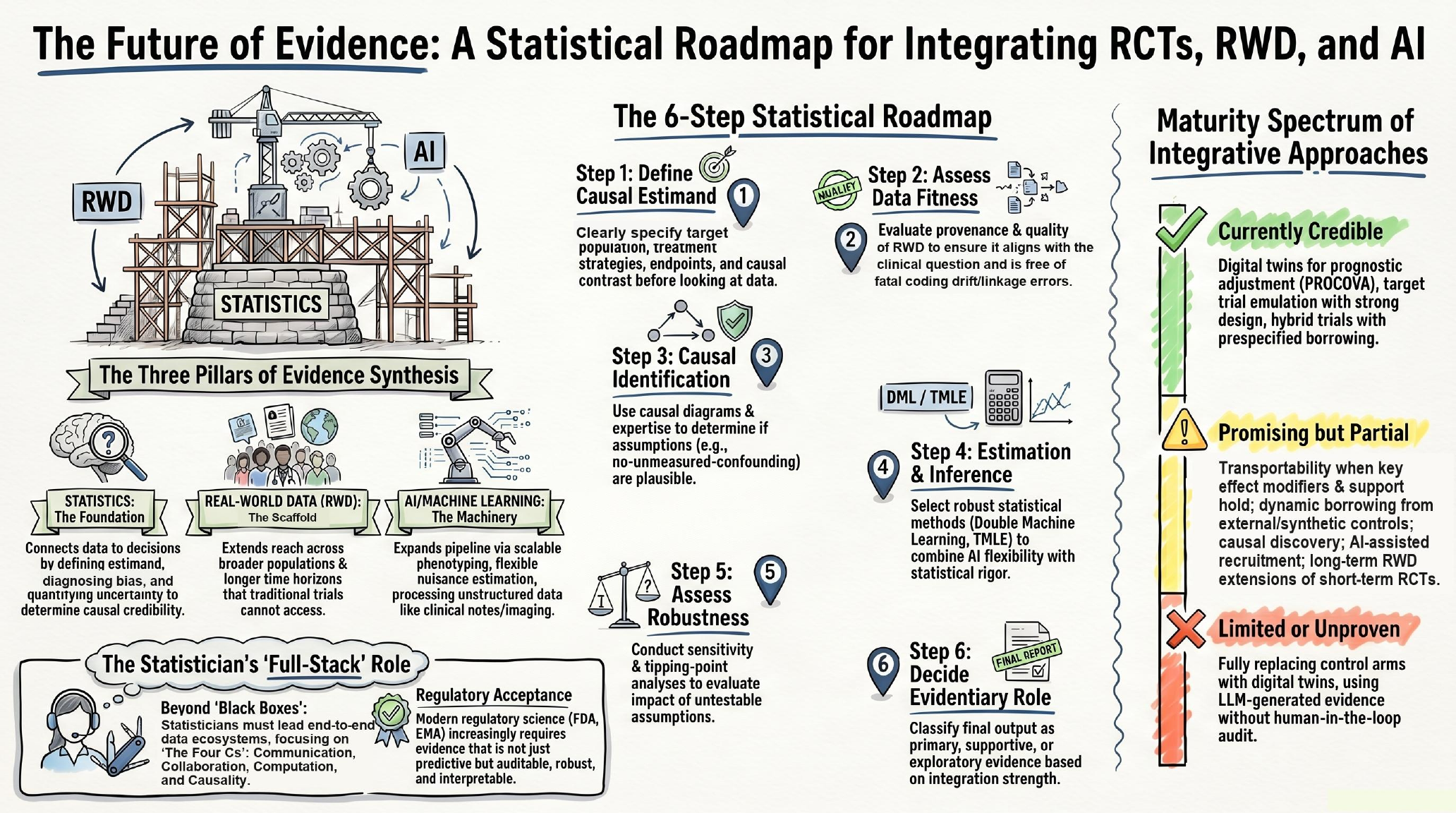}
\vspace{1.2em}
\caption{A statistical evidence roadmap for integrative evidence synthesis.
(Left) Complementary strengths of RCTs, RWD, and AI/ML as evidence
sources and the statistician's full-stack role. (Middle) The six-step
roadmap spanning design, analysis, and decision phases. (Right) Key
applications of RCT--RWD integration, annotated by regulatory readiness.}
\label{fig:roadmap}
\end{figure}

\paragraph*{AI/ML across the evidence lifecycle.}
AI/ML methods are rapidly reshaping RWE generation
and biomedical research, offering capabilities that classical analytical
tools cannot match across a range of data types, including unstructured
clinical notes, medical imaging, wearable sensor streams, and graph-structured
biological networks, while also enabling automation at multiple stages
of the evidence pipeline.

\textit{Upstream}, AI can support
systematic literature retrieval and structured extraction of published
findings to inform study design. Large language models can identify
endpoints and operationalize phenotypes directly from clinical notes;
computer vision methods can quantify imaging features at scale; and
multimodal representation learning can enrich covariate characterization
by integrating heterogeneous data sources. These upstream tasks shape
how treatments, outcomes, and covariates are defined, and therefore
influence the treatment-effect estimand itself, often before statistical
estimation begins. In clinical trial operations, agentic AI systems
have been proposed to support adaptive recruitment optimization aimed
at improving participant diversity \citep{harrer2019artificial} and
real-time operational adjustments in decentralized trial settings
\citep{thomas2022artificial}. \textit{In the middle},
AI and ML methods offer flexibility where conventional approaches
are too rigid: high-dimensional propensity score and outcome modeling
with text and image inputs, estimation of treatment-effect heterogeneity,
and the construction of synthetic controls and digital twins, among
other applications. \textit{Downstream},
AI agents can automate reporting and result archiving, monitor model
drift, and maintain living evidence pipelines that update analyses
continuously as new data accumulate, which is particularly relevant
in post-market settings where evidence generation is ongoing rather
than episodic.

However, predictive power and automation alone do
not confer causal validity or regulatory credibility. The central
methodological challenge remains unchanged: determining which conclusions
are scientifically interpretable, robust to sources of bias, and sufficiently
trustworthy to support consequential decisions.

\paragraph*{Statistics remains indispensable.}
Statistics is the discipline that transforms data
into reliable knowledge by accounting for uncertainty, variability,
bias, and context. It provides the principles, methods, and tools
for study design, data collection, variation modeling, uncertainty
quantification, inference, and decision support. For treatment-effect
evaluation, this means attending to the validity of identifying assumptions,
coverage properties of confidence intervals, and error-rate control. AI/ML methods are often strongest for pattern recognition,
data compression, multimodal feature construction, flexible nuisance
estimation, and workflow automation, while statistical methods remain
central when the goal is causal effect estimation, uncertainty quantification,
evidence synthesis, and transparent decision support.

In the integrated landscape of RCTs, RWD, and AI/ML,
the role of statistics is not diminished but broadened, and its relationship
to the other components can be understood through a structural metaphor
(Figure~\ref{fig:roadmap} Left). Statistics is the \textit{foundation}:
it defines the scientific question, specifies the estimand, diagnoses
threats to identification, quantifies uncertainty, and determines
what is causally credible and decision-ready. Real-world data is the
\textit{scaffold}: it extends
the reach of evidence across broader populations, longer time horizons,
and real-world settings that trials cannot access alone. AI is the
\textit{machinery}: it expands
what can be measured, linked, represented, and automated across the
evidence pipeline. Each component depends on the others, but the foundation
is what determines whether the structure holds. This framing aligns
with recent calls for statisticians to engage more actively in end-to-end
data ecosystems and to communicate the field's distinctive contributions
in bias assessment, uncertainty quantification, and causal reasoning
\citep{donoho2026rebuilding}.

We therefore argue that the central challenge is
neither RCTs versus RWD nor statistics versus AI/ML, but the principled
integration of all three across the full evidence lifecycle from discovery
and study planning through design, analysis, post-market surveillance,
and consequential decision-making.

\paragraph*{Scope and core questions.}
This Perspective is written from that statistical
viewpoint. Rather than providing a comprehensive review of all recent
developments, we focus on a set of core questions that statisticians
must confront:  
\begin{enumerate}
\item[(Q1)] When is RWD fit for purpose for causal and regulatory
use, and when should statisticians say no?  
\item[(Q2)] What does AI genuinely add across the evidence lifecycle
that conventional statistical approaches alone cannot achieve at scale?  
\item[(Q3)] What remains distinctly statistical and indispensable
even in an AI-rich workflow?  
\item[(Q4)] How should statisticians build trustworthy end-to-end
evidence systems in pharma, industry, and regulatory science?  
\item[(Q5)] How should statistical training evolve so that statisticians
can lead this space?  
\end{enumerate}
A central theme is that integration must be governed
by a statistical evidence roadmap. RWD is not automatically fit for
causal use simply because it is large, and AI/ML does not rescue poor
study design or unresolved bias. Key threats such as unmeasured confounding,
informative observation processes, temporal misalignment, endpoint
misclassification, coding drift, linkage error, and selection bias
are not minor caveats but central constraints. Some can be partially
mitigated through principled design, causal inference, calibration,
validation studies, or sensitivity analysis. Others may render a causal
estimand insufficiently credible for the intended decision. A useful
framework must therefore help determine when integration is scientifically
persuasive, when it is only partially reliable, and when one should
stop.

The remainder of the paper is organized around this
vision. Section~\ref{sec:roadmap} introduces a statistical evidence
roadmap for integrative evidence generation (Figure~\ref{fig:roadmap}
Middle). Section~\ref{sec:applications} presents applications of
the roadmap across several integrative scenarios (Figure~\ref{fig:roadmap}
Right). Section~\ref{sec:future} presents a forward-looking agenda
for what statisticians should build, learn, and lead next.

\section{A Statistical Evidence Roadmap for Integrative Evidence Generation}

\label{sec:roadmap}

The future of evidence generation will not be determined
by choosing between RCTs, RWD, conventional statistics, and AI/ML.
It will be determined by how effectively these components are integrated
to answer questions of clinical and regulatory importance with sufficient
transparency, robustness, and credibility.

\subsection{Why a statistical evidence roadmap is needed}

\label{sec:roadmap-why}

This subsection directly addresses \textbf{(Q1)}:
when is RWD fit for purpose for causal and regulatory use, and when
should statisticians say no?

\paragraph*{FDA guidance: bias concerns with RWD.}
Real-world data sources may introduce biases that, if unaddressed,
threaten the validity of study conclusions. These include selection
bias, when RWD populations differ systematically from those in RCTs;
unmeasured confounding, when key clinical variables are missing or
inconsistently captured; immortal time bias, arising from temporal
misalignment when treatment initiation and outcome follow-up are not
anchored to a common baseline, an issue that can be exacerbated when
RWD and RCT data originate from different care settings; and measurement
error, when outcomes recorded in routine care do not align with trial-defined
endpoints. These are not minor details but fundamental threats to
causal validity and regulatory confidence \citep{FDA2018}.

\paragraph*{Limitations of AI/ML and the enduring value of statistics.}
Many AI/ML models have been criticized as ``black
boxes,'' characterized by limited interpretability, inadequate uncertainty quantification, and susceptibility to bias when applied to non-representative data. In contrast, statistics
emphasizes inference and explanation, yielding interpretable estimates
grounded in explicit assumptions. Statistical methods provide well-established
tools for uncertainty quantification (e.g., confidence intervals,
hypothesis testing, and p-values) and reproducibility, which remain central
to regulatory credibility \citep{cox1958planning}.

\paragraph*{Where we go next.}
Each component brings a distinct strength: RCTs secure internal validity;
RWD/E extends generalizability; statistics anchors validity, interpretability,
and regulatory acceptance; AI/ML expands the analytical horizon to
complex, large-scale datasets and novel data sources. What is needed
is a statistical evidence roadmap: a unifying framework that makes
estimands and assumptions explicit, clarifies trade-offs, and preserves
scientific rigor while maximizing practical relevance.

\subsection{The six-step roadmap for evidence integration}

Integrating RCTs with RWD requires data that are fit for purpose and
a principled roadmap grounded in causal inference. At the foundation
lies trust in data quality, provenance, and transparency. How data
are generated, curated, and validated for specific causal questions
determines whether the resulting evidence will be credible to regulators,
clinicians, and patients alike.

Building on this agenda, we present a statistical evidence roadmap for integrating
RCTs, RWD, and AI/ML across the evidence lifecycle (Figure~\ref{fig:roadmap}
Middle). This roadmap builds on the Causal Roadmap for generating high-quality real-world evidence \citep{dang2023causal}, which organizes estimand definition, characterization of observed data and data-generating processes, identification, estimation, and sensitivity analysis. Our contribution extends that framework in three respects: (i) explicit integration of AI/ML across the evidence lifecycle, rather than treating flexible learning only as an estimation device; (ii) a maturity classification that distinguishes currently credible, partially reliable, and unproven integrative applications; and (iii) an explicit final step assigning the evidentiary role of the integrated analysis (primary, supportive, exploratory, descriptive, or unsuitable). The roadmap is intended to be operational: it identifies
where evidence generation is on solid ground, partially reliable,
or should stop. 
\begin{enumerate}
\item[\textbf{Step$\ $1.}] \textbf{Define the decision problem and causal estimand.} The
starting point is not the data source or the algorithm, but the question
to be answered. One must specify the target population, treatment
strategies, endpoint, intercurrent event handling, follow-up horizon,
and causal contrast of interest. This step is foundational because
many downstream problems arise when the scientific question is not
clearly distinguished from what the available data happen to support.
In integrative settings, the estimand also clarifies the intended
role of each data source: whether RWD is being used to broaden the
target population, extend follow-up, augment/supplement a control
arm, improve efficiency, or support post-market surveillance. 
\item[\textbf{Step$\ $2.}] \textbf{Assess data provenance, alignment, and fitness
for purpose.} Understand where data came from,
how they were collected, and whether they contain the right information
at sufficient quality to address the causal question. In
integrative settings, this is often the most consequential step. Incompatible
baseline definitions, coding drift, linkage error, treatment and outcome
misclassification, and informative observation processes can silently
bias and invalidate causal conclusions. AI can help at this stage
through phenotyping, semantic alignment, multimodal extraction, and
workflow support, but these tools do not eliminate the need for scientific
judgment.  
\item[\textbf{Step$\ $3.}] \textbf{Determine whether causal identification
is plausible.} This step asks whether the estimand
can, in principle, be linked to observed data under assumptions that
are scientifically credible. No unmeasured confounding, for example,
requires that all variables influencing both treatment assignment
and outcome are measured and adequately captured. Causal diagrams
\citep{greenland1999causal} informed by subject-matter expertise
remain useful for interrogating these assumptions, and in modern high-dimensional
settings they may be complemented by causal discovery algorithms,
biomedical knowledge graphs, and structured expert-in-the-loop variable
selection. Positivity, or common support, should likewise be assessed explicitly: within confounder strata, each treatment strategy of interest must have positive probability, and analogous participation or selection positivity is required for target-trial emulation, external-control borrowing, heterogeneous treatment effect (HTE) estimation, and transportability. Limited overlap can be diagnosed via extreme weights and effective sample size, and addressed by trimming, restricting to a common-support subpopulation, avoiding unsupported extrapolation, or redefining the target population or estimand. Such discovery tools can scale hypothesis generation but do not
replace scientific reasoning or identification theory. The roadmap
should explicitly distinguish between biases that may be mitigable
and those that may be intrinsic. If key confounders are systematically
absent, if positivity fails in regions needed for the estimand, if treatment initiation cannot be anchored to a meaningful
time zero, or if the observation process is strongly informative in
ways that cannot be modeled or stress-tested, then the appropriate
conclusion may be that causal identification is not sufficiently credible
for the intended use. 
\item[\textbf{Step$\ $4.}] \textbf{Select estimation and inference methods appropriate to the
estimand.} Statistical and AI/ML tools should be chosen to align with
the causal estimand and the identification assumptions invoked. Flexible
ML algorithms are well-suited to nuisance function estimation in high-dimensional
settings, but can remain biased in finite samples when used naively.
Doubly robust estimators and double/debiased machine learning (DML) are related but distinct. Classical double robustness yields consistency if either the propensity score or the outcome regression is correctly specified \citep{bang2005doubly}; DML combines Neyman-orthogonal scores with cross-fitting to support valid inference when nuisance functions are estimated by flexible ML under product-rate and regularity conditions \citep{chernozhukov2018double}. In the DML setting, valid inference for a fixed-dimensional treatment-effect parameter can be retained when nuisance estimators converge more slowly than $n^{-1/2}$, provided that the product of nuisance errors is $o_p(n^{-1/2})$ and appropriate cross-fitting is used. Multimodal
representations drawn from clinical notes, patient narratives, and
embeddings can further enrich covariate construction in complex data
environments. In small-sample or randomized settings, randomization-based
inference \citep{fisher1935} may offer stronger finite-sample operating
characteristics than asymptotic alternatives. 
\item[\textbf{Step$\ $5.}] \textbf{Assess robustness.} Uncertainty quantification and sensitivity
analysis are central inferential components. In integrative settings,
important assumptions are often partially or entirely untestable,
especially when combining external data with randomized evidence or
when using AI-derived variables and models. Robustness assessment
should be structured and targeted: this may include bias functions,
calibration analyses, alternative endpoint definitions, tipping-point
analyses, or prospective validation studies. 
\item[\textbf{Step$\ $6.}] \textbf{Decide the evidentiary role of the integrated
analysis.} Not every integrated analysis should
be treated as primary causal evidence. A final step should classify
the role of the resulting evidence as primary, supportive, exploratory,
descriptive, or unsuitable for causal interpretation. This is especially
important for regulatory and industry settings, where the same analytic
pipeline may be used for very different decision contexts. Making
this classification explicit helps prevent overstatement and keeps
the interpretation aligned with the actual strength of the evidence. 
\end{enumerate}

\paragraph*{Implications for statisticians.}
This roadmap implies a broader, \textquotedbl full-stack\textquotedbl{}
role for statisticians in modern evidence generation: one that spans
target trial specification, data curation and provenance assessment,
causal reasoning, method selection, validation and sensitivity design,
through to interpretation, communication, and the shaping of standards
by which AI-assisted evidence pipelines are judged.

\section{Applications of the Statistical Evidence Roadmap }

\label{sec:applications}

In regulatory science, integrative evidence synthesis is not about
replacing RCTs with RWD, but about filling evidence gaps that RCTs
alone cannot address. The goal is to determine when and how RWD can
reliably complement or extend RCT findings to support regulatory,
clinical, and reimbursement decisions. Achieving this requires data
that are demonstrably fit for purpose and integration governed
by the statistical evidence roadmap introduced in Section~\ref{sec:roadmap}.
The applications below collectively address \textbf{(Q1)--(Q3)}:
when RWD is fit for purpose, what AI/ML genuinely adds, and what statistical
rigor remains indispensable. 

A common thread across all applications below is that RWD and AI/ML
contribute genuine value, but only when embedded within a principled
statistical framework. Without the roadmap, RWD- and AI/ML-assisted
analyses risk amplifying biases, conflating prediction with causal
inference, or producing impressive-looking results that lack scientific
credibility. With the roadmap, the same RWD and AI/ML tools become
powerful components of a trustworthy evidence system. Each application
below illustrates this principle by showing how the six-step roadmap
disciplines the use of RWD and AI/ML at every stage,
from defining the estimand through classifying the evidentiary role
of the resulting analysis.

\subsection{AI-assisted recruitment through active learning}

AI-assisted trial design introduces an active learning paradigm in
which RWD and AI/ML identify participants most informative for the
trial's scientific aims. When the objective is to estimate an optimal
treatment decision rule, selectively recruiting patients near the
decision boundary improves information efficiency. When the goal is
to evaluate treatment efficacy, prioritizing individuals with the
largest expected treatment effects increases statistical power, reduces
the required sample size, and is more ethical by
providing faster access to promising treatments for relevant patients.

\paragraph*{Applying the roadmap.}
Active learning requires special attention at Steps 1 and 6. The estimand
(Step 1) must specify the intended target population, not merely the
enriched subpopulation selected by the algorithm. Data fitness (Step
2) must evaluate whether the RWD guiding recruitment is itself representative,
since AI/ML trained on biased RWD may systematically enrich subpopulations
correlated with access to care. Identification (Step 3) should assess
whether enrichment alters the effective target population in ways
that compromise external validity. Moreover, one can
improve external validity through generalizability and transportability
analysis, as discussed in Section~\ref{subsec:Generalizing-and-transporting},
which reweights the trial population toward a target
population so that conclusions better reflect real-world practice.
The evidentiary role (Step 6) must be explicit: AI-assisted recruitment
involves a trade-off between efficiency and representativeness, and
resulting treatment effect estimates should be interpreted as conditional
on the enriched mechanism unless transportability to the intended
population can be justified.

\subsection{AI-assisted RCT analysis: super-covariate adjustment}

AI can strengthen the analysis of RCTs without compromising the validity
conferred by randomization. Among generative AI applications leveraging
RWD, digital twins, namely virtual, patient-specific comparators that
simulate disease trajectories, have become one of the most mature
paradigms. PROCOVA (Prognostic Covariate Adjustment), developed by
Unlearn.AI, has been piloted in Alzheimer's disease and amyotrophic
lateral sclerosis trials, enabling reductions in control arm sizes
while maintaining statistical power \citep{schuler2021procova,ross2024enhancing}.
Randomization guarantees internal validity; ANCOVA
improves efficiency by adjusting for prognostic covariates. PROCOVA
extends this by incorporating digital twin predictions as a \textquotedbl super-covariate\textquotedbl{},
producing more precise estimates and enabling
smaller sample sizes. Importantly, the method does not require AI
predictions to be correct: it benefits when they are, but the protection
of randomization is preserved regardless.

\paragraph*{Applying the roadmap.}
The estimand (Step 1) remains the average treatment effect in the
RCT population, protected by randomization. AI enters only through
Step 4 as an efficiency tool, not to define the scientific question.
Data fitness (Step 2) requires validating that RWD used to train the
digital twin aligns with the trial population to yield meaningful
efficiency gains. Identification (Step 3) is anchored by randomization,
so concerns shift to whether the AI-derived covariate introduces instability
or hidden dependencies. Robustness (Step 5) should evaluate sensitivity
to digital twin prediction quality. The evidentiary role (Step 6)
remains primary: the randomized design provides the causal anchor,
and AI improves precision without altering the inferential target.

More broadly, integrating AI-derived covariates with doubly robust
estimators such as Augmented Inverse Probability Weighting, Targeted
Maximum Likelihood Estimation, or Double Score Matching \citep{tan2025double}
allows statisticians to exploit complex nonlinear relationships while
retaining $\sqrt{n}$ consistency and asymptotic normality under standard semiparametric conditions on nuisance estimation rates and regularity; these properties are not automatic for arbitrary machine-learning plug-in estimators.

\subsection{Target trial emulation}

Target trial emulation has emerged as a principled framework for using
observational data to approximate randomized conditions, extending
evidence to new indications, prolonging follow-up beyond trial windows,
and addressing long-term safety and effectiveness questions \citep{hernan2016using}.
The approach requires explicit specification of a hypothetical trial
protocol: eligibility criteria, treatment strategies, assignment procedures,
follow-up period, outcome definition, causal contrast, and analysis
plan.

\paragraph*{Applying the roadmap.}
The estimand (Step 1) is defined by the hypothetical target trial
the observational analysis seeks to emulate. Related approaches such
as \textit{Trial Pathfinder} \citep{liu2021evaluating} offer data-driven
optimization of eligibility criteria using RWD, extending the framework
to settings where the target population itself is under refinement.
Data fitness (Step 2) is critical: RWD must faithfully reconstruct
eligibility, treatment initiation, and follow-up structure. Identification
(Step 3) hinges on whether the emulated design adequately addresses
confounding by indication, immortal time bias, and informative observation
processes; each represents a distinct threat to validity that causal
diagrams and sensitivity analyses should systematically probe. Electronic
health records are particularly susceptible to informative observation:
when the timing and frequency of measurement depend on patient status
or clinical trajectory, analyses that treat the observation process
as ignorable can introduce substantial bias and require explicit modeling.
The evidentiary role (Step 6) must reflect an irreducible limitation:
emulation approximates but does not replicate randomization. Results
are therefore best classified as supportive for established indications
or hypothesis-generating for new ones, and should be interpreted accordingly.

\subsection[Synthetic controls and generative AI]{Synthetic controls and generative AI}

Recent work has explored generative AI for constructing synthetic
control cohorts, particularly in oncology where enrollment to control
arms competes directly with access to investigational therapy \citep{eckardt2026artificial}.
Proof-of-concept studies have demonstrated that AI-generated synthetic
patients can replicate survival outcomes from completed trials in
acute myeloid leukemia, colorectal cancer, and non-small-cell lung
cancer.

Synthetic controls, however, carry risks that are distinct from and
additive to those of conventional external controls. The quality of
synthetic data is bounded by the representativeness of the training
cohort, which is a circular dependency that more sophisticated architectures
do not resolve. More fundamentally, standard fidelity metrics such
as distributional similarity and train-on-synthetic-test-on-real performance
do not assess whether causal structure is preserved. A synthetic cohort
may closely match marginal distributions while failing to reproduce
confounding patterns, informative censoring mechanisms, or time-varying
treatment effects that are essential to valid causal inference.

\paragraph*{Applying the roadmap.}
The estimand (Step 1) must specify what population the synthetic controls
are intended to represent and whether the causal contrast remains
interpretable when one arm is entirely generated. Data fitness (Step
2) requires that the training cohort reflect current standard of care
and cover relevant subpopulations, with attention to coding consistency
and temporal alignment. Causal identification (Step 3) is fundamentally
constrained: synthetic generation preserves associations but not necessarily
the data-generating process, and unmeasured confounding in the training
data propagates directly into the synthetic output. Estimation methods
(Step 4) must account for the additional uncertainty introduced by
the generative process itself, beyond what conventional variance estimation
captures. Robustness assessment (Step 5) should examine sensitivity
to generative model choice, hyperparameter settings, and training
sample size; multiple studies report that performance deteriorates
with longer follow-up and smaller training sets. The evidentiary role
(Step 6) should currently be classified as exploratory or hypothesis-generating
rather than primary or supportive, consistent with the field's present
maturity.

\subsection{Hybrid controlled trials: borrowing external information with statistical
safeguards}

Replacing control arms entirely with real-world comparators or AI-derived
digital twins is risky, but hybrid designs offer a more reliable alternative.
These designs retain a smaller randomized control arm to anchor inference
while borrowing information from digital twins and external controls
in a bias-aware manner. The causal estimand remains the average treatment
effect in the RCT population, while concurrent controls enable bias
detection and calibration of external sources.

Three principles govern this integration. First, randomization remains
the gold standard: hybrid trials benefit from the internal benchmark
provided by the RCT for unbiased estimation and for diagnosing discrepancies
in external controls. Borrowing methods cannot compensate for the
absence of randomization when unmeasured confounding is substantial.
Second, sensitivity analysis should be treated as a core inferential
component, since key assumptions linking external data to the trial
population are untestable. Third, careful trial design is essential:
design choices such as unbalanced randomization intended to facilitate
borrowing may reduce efficiency if external information ultimately
contributes little.

\paragraph*{A continuum of Bayesian and frequentist methods.}
Bayesian approaches adaptively down-weight or filter external data
when discrepancies arise, using power priors, commensurate priors,
meta-analytic predictive priors, and hierarchical exchangeability
models \citep{chen2000power,hobbs2011hierarchical,schmidli2014robust,Alt2024}.
Frequentist approaches include test-then-pool rules \citep{viele2014use},
bias-function modeling \citep{van2024adaptive}, study-level selection
\citep{chen2021combining}, and semiparametric selective borrowing
\citep{gao2024doubly,gao2025improving} that reweights or borrows
only comparable subsets. In rare disease trials, Fisher's randomization
test \citep{zhu2025enhancing} provides an assumption-light safeguard
with exact type-I error control even in small samples. Together, these
strategies form a continuum from blunt screening to precise adjustment
to definitive testing.

\paragraph*{Applying the roadmap.}
The estimand (Step 1) is defined within the RCT population. Data fitness
(Step 2) requires rigorous assessment of whether external controls
are comparable in eligibility, endpoints, and follow-up. Identification
(Step 3) depends on the plausibility that external and internal controls
are conditionally exchangeable, a strong assumption requiring stress-testing.
Method selection (Step 4) should match borrowing degree to confidence
in compatibility. Robustness (Step 5) is non-negotiable. The evidentiary
role (Step 6) should reflect that hybrid designs produce primary evidence
only when the randomized component is large enough to anchor inference;
otherwise, borrowing is supportive.

\subsection{RWD-augmented RCTs: heterogeneous treatment effects}

\label{subsec:HTE}Characterizing how treatment effect
varies across patient characteristics (i.e., HTE) is central to precision medicine and personalized care.
HTE estimation is arguably where the combined power of RCTs, RWD,
and AI/ML is greatest: RCTs provide unbiased anchoring, RWD supplies
rich patient heterogeneity, and AI/ML enables flexible modeling of
complex interactions without restrictive parametric assumptions; yet
this opportunity remains underweighted in integrative evidence frameworks.

\paragraph*{Applying the roadmap.}
The estimand (Step 1) specifies the conditional treatment
effect as a function of effect-modifying covariates, anchored on the
RCT population; extending to a broader target population may require
additional transportability assumptions (Section~\ref{subsec:Generalizing-and-transporting}).
Data fitness and identification (Steps 2--3) require that both RCT
and RWD capture key effect modifiers at sufficient quality, with conditional
exchangeability holding within covariate strata. Estimation (Step
4) employs approaches robust to identification violations, including
elastic integrative estimators \citep{yang2023elastic}, bias-function
modeling \citep{wu2022integrative,yang2024datafusion}, and harmonization
\citep{schwartz2026harmonized}. Robustness (Step 5) examines sensitivity
to the assumed effect-modifier set and subgroup stability under covariate
shifts. The evidentiary role (Step 6) is typically exploratory unless
prospectively pre-specified and validated.

\subsection{Extending short-term RCTs with long-term RWD}

Integrating short-term RCTs with long-term RWD provides a powerful
framework for evaluating treatment durability, late safety signals,
and downstream outcomes that RCTs alone cannot capture due to limited
follow-up. Traditional trials often stop at intermediate
endpoints or relatively short time horizons, leaving critical questions
about sustained benefit and the implications of extending the trial period
unanswered. For example, hybrid analyses linking oncology RCTs with
cancer registries or electronic health records have been used to extend
survival follow-up, assess recurrence rates, and monitor late-onset
or long-latency adverse events. Such integration has been applied
in immuno-oncology and cardiovascular studies, where trial evidence
demonstrates early efficacy but real-world registries reveal longer-term
patterns of relapse, toxicity, or adherence. By leveraging RWD to
extend the evidentiary window, researchers can generate evidence on
durability that directly informs clinical guidelines, regulatory decisions,
and payer assessments of value.

\paragraph*{Applying the roadmap.}
The estimand (Step 1) must specify long-term survival, effectiveness,
or safety in the relevant target population. Data fitness (Step 2)
requires careful alignment of endpoints, coding systems, and follow-up
definitions. Identification (Step 3) must address informative censoring,
treatment switching, and unmeasured confounding that accumulate over
extended follow-up. Methods (Step 4) include bias-function modeling
\citep{zhou2024causal}, surrogate index \citep{athey2025surrogate},
and data fusion strategies \citep{imbens2025long}. The evidentiary
role (Step 6) is typically supportive: extended analyses inform
such decisions but do not
replace primary trial evidence.

Health technology assessment agencies have increasingly highlighted
the importance of such strategies, particularly in oncology and rare
diseases where long-term randomized evidence is impractical.

\subsection{Generalizing and transporting RCT findings to broader populations}

\label{subsec:Generalizing-and-transporting}

Integrating RCTs with RWD enables evidence that extends trial findings
to more diverse and representative patient populations. In RCTs, strict
eligibility criteria often limit representativeness: patients with
comorbidities, polypharmacy, or advanced disease are typically underrepresented.
While these restrictions ensure internal validity and safety, the
resulting trial populations may not reflect how treatments perform once approved, where physicians
exercise clinical judgment in tailoring therapies, adjusting treatment
sequencing, and managing concurrent conditions.

Illustrative applications include generalizing treatment effects from
a lung cancer trial to the National Cancer Data Base, revealing that
trial participants were typically younger and healthier than real-world
patients, and transporting survival effects from an HIV trial to populations
in the U.S., Thailand, and Ethiopia \citep{lee2022doubly,lee2024transporting}.
Beyond the FDA, health technology assessment agencies such as Germany's
G-BA (2020), the U.K.'s NICE (2013), and the U.S. ICER (2020) explicitly
emphasize evidence addressing generalizability and representativeness
to guide reimbursement and market access decisions.

\paragraph*{Applying the roadmap.}
Generalization and transportability follow the six steps directly.
Define the causal estimand in the target population (Step
1), using population-based registries, national health surveys, or multinational
trial networks. Assess data fitness (Step 2) by examining
whether the RWD captures the key effect-modifying covariates needed
for transportability, with attention to endpoint harmonization and
coding consistency. Evaluate identification (Step
3) by articulating assumptions regarding internal validity of the
RCT and transportability; e.g., causal diagrams clarify identification
and diagnose threats. Transportability requires that the treatment
effect function behaves similarly across populations, hinging on alignment
of key effect-modifying covariates. It also requires trial-participation or selection positivity: regions of the covariate space with negligible overlap between trial participants and the target population should be diagnosed through weight diagnostics and effective sample size, with remedies such as restricting extrapolation or redefining the target estimand. Select methods (Step
4) such as doubly robust estimators and augmented calibration weighting
\citep{lee2023improving} that combine flexible nuisance estimation
with rigorous inference; here, AI/ML strengthens nuisance function
estimation without compromising inferential validity. Conduct sensitivity
analyses (Step 5); recent work \citep{jin2024beyond}
demonstrates that unobserved conditional shifts often dominate over
covariate shifts. Classify the evidentiary role (Step
6): transported estimates typically serve as supportive evidence
rather than primary proof of efficacy.

Transportability is closely linked to reproducibility.
When populations are sufficiently similar, treatment effects estimated
in one population should replicate in another. Ensuring this reproducibility
not only corroborates causal assumptions but also reinforces confidence
in the generalizability and trustworthiness of treatment discoveries.

\subsection{Where Integration Succeeds, Partially Succeeds, and Remains Unproven}

\label{sec:future-maturity}

As illustrated in Figure~\ref{fig:roadmap} Right,
the maturity of integrative approaches varies considerably.

\paragraph*{Currently credible:}
AI-assisted prognostic covariate adjustment within
randomized trials; target trial emulation in settings with strong
design alignment; hybrid or RWD-augmented trials with careful prespecification.

\paragraph*{Promising but partial:}
Transportability and generalizability, currently more credible when key effect modifiers are measured and positivity/common support holds, but not when modifier capture or overlap is incomplete;
dynamic borrowing from external or synthetic controls
with unanchored comparisons; causal discovery in high-dimensional
clinical systems; AI-assisted recruitment or enrichment; long-term RWD extension of short-term RCTs.

\paragraph*{Limited or unproven:}
Fully replacing controls with digital twins or synthetic
cohorts in high-stakes settings; causal inference when key confounders
are fundamentally unmeasured; generative AI-based evidence generation
without auditability and hallucination control; synthetic data for
primary regulatory evidence without independent validation of causal
structure preservation.

These applications demonstrate that integrative methods
are no longer theoretical: they are actively shaping regulatory and
clinical decision-making. Across all of them, the
principle is consistent: AI expands what can
be measured, modeled, and automated; statistics determines what is
causally credible, interpretable, and decision-ready.

\section{What Statisticians Should Build Next and Learn }

\label{sec:future}

\subsection{A Forward-Looking Agenda}

The field of evidence generation stands at a pivotal juncture. The
confluence of increasing data availability, rapid advances in AI/ML,
and pressing demands from regulators, sponsors, payers,
providers, and patients presents both challenges and opportunities.
Issues such as uncertainty, small-sample inference, privacy, and bias
must be addressed systematically, but they should be viewed as catalysts
for innovation rather than unsolvable obstacles. Below we outline
key open methodological problems representing natural opportunities
for statistical leadership. 

\paragraph*{Uncertainty quantification in AI-rich workflows.}
As evidence increasingly combines RCTs, RWD, and AI/ML predictions,
reliable uncertainty quantification is essential. \textit{Conformal
prediction} offers distribution-free, finite-sample valid predictive
inference for black-box models under exchangeability \citep{vovk2005algorithmic},
with extensions to individualized treatment effects \citep{lei2021conformal}
and joint efficacy--safety evaluation \citep{jin2023sensitivity} as
active frontiers.
A related challenge is inference for AI-derived covariates and endpoints.
When phenotypes are extracted by LLMs or imaging biomarkers are quantified
by deep networks, valid frameworks for propagating measurement uncertainty
through downstream causal estimands are needed. This combines errors-in-variables
methodology, post-prediction inference, and semiparametric efficiency
theory.
Calibrated synthetic-control inference presents a third open problem.
Standard inferential procedures assume a fixed data-generating process;
synthetic data introduce an additional layer of stochastic variation
from the generative model itself, and methods for propagating that
uncertainty through downstream causal analyses remain underdeveloped.
This is a critical gap: regulatory applications require not only point
estimates but calibrated measures of evidential strength. A promising
direction is to adapt \textit{multiple imputation}
frameworks, originally developed for missing data, to the high-dimensional,
nonlinear mappings of modern generative models, treating synthetic
generation as a structured source of inferential uncertainty rather
than a resolved preprocessing step.

\paragraph*{Small-sample challenges.}
Many high-stakes clinical contexts, including rare diseases and early-phase
drug development, operate under severe data constraints. Traditional
asymptotic approximations often break down, and naive machine learning
algorithms may overfit, yielding unstable or irreproducible estimates.
Addressing this requires algorithms and estimators specifically designed
for finite samples: approaches that balance efficiency with robustness,
make optimal use of external or historical controls, and provide valid
inference under realistic assumptions. For instance, randomization-based
inference frameworks can help ensure that findings remain credible even
in data-limited scenarios. Such methods not only expand the feasibility
of trials in ultra-rare conditions but also increase equity by enabling
evidence generation in historically underserved populations. Emerging
developments, such as conformal inference, can strengthen finite-sample
predictive coverage guarantees and calibrate uncertainty across RCT
and RWD sources.

\paragraph*{Privacy-preserving evidence generation.}
Protecting patient privacy while enabling meaningful multi-institutional
research, an issue that also carries geopolitical
implications, remains one of the most urgent concerns in modern data
integration. \textit{Federated learning} offers a promising solution
by allowing decentralized analyses across hospitals, registries, and
trial networks without sharing individual-level data \citep{li2020federated}.
Recent developments in federated causal inference, such as adaptive
weighting schemes, penalized regression frameworks, and communication-efficient
protocols, make it possible to estimate causal effects across diverse
populations while minimizing privacy risks. These methods are especially
valuable in rare diseases and underrepresented populations, where
single-institution data are insufficient. Beyond methodology, widespread
adoption will require interoperable infrastructures, secure platforms,
and regulatory clarity on how federated analyses can be incorporated
into regulatory submissions and health technology assessments.

Generative AI offers a complementary path. By learning
the underlying data distribution rather than memorizing individual
records, generative models can synthesize realistic, non-identifiable
datasets that retain essential statistical properties of the source
data. However, privacy is not guaranteed. Some generative methods,
particularly diffusion-based frameworks, may support privacy-aware
data sharing or simulation under carefully validated conditions, but
preservation of clinically and causally relevant structure must be
demonstrated rather than assumed. Different generative paradigms have
different risks, and claims about privacy-preserving analytics cannot
be transferred indiscriminately across all generative and language-based
models. Membership-inference and model-inversion attacks can re-identify
individuals from synthetic outputs, particularly for rare subgroups
\citep{yang2025model}. Training data biases may be amplified rather
than mitigated, and differential privacy introduces a fidelity-privacy
tradeoff that can distort the distributional properties needed for
valid inference. A dual-layered architecture combining federated learning
with carefully validated generative modeling remains a promising long-term
direction, but its promise is contingent on resolving these empirical
challenges first.

\paragraph*{Prospective validation of causal assumptions in
RWE.}
Evaluating the causal assumptions that underpin
real-world evidence is critical, yet many such assumptions cannot
be fully verified using retrospective observational data alone. A
practical and scalable strategy is to embed small, prospective randomized
studies designed specifically to assess the robustness of these assumptions.
For example, consider a situation in which a large RWD analysis indicates
that treatment A is more effective than treatment B in reducing cardiovascular
risk. While a full-scale RCT may be impractical due to cost or operational
constraints, a smaller randomized validation study can still yield
meaningful evidence regarding the credibility of this causal conclusion.
One approach is to use estimated propensity scores from the RWD to
inform randomization probabilities in the validation study. By randomizing
patients conditional on their covariate-adjusted scores, confounding
can be eliminated by design. The treatment effects observed in this
randomized validation cohort can then be used to quantify, in probabilistic
terms, the degree to which the RWD-based conclusion about treatment
superiority is likely to hold. This hybrid strategy strengthens causal
claims while maintaining feasibility and efficiency.

\paragraph*{Causal inference under model and data drift.}
AI-assisted pipelines, post-market surveillance, and
living evidence systems all operate under conditions of non-stationarity:
patient populations shift, coding practices evolve, clinical guidelines
change, and AI model behavior degrades over time. Developing sequential
testing procedures, online inference frameworks, and drift-robust
estimators that maintain valid statistical guarantees as the data-generating
process evolves is a natural area of statistical leadership. Such
methods would directly support the living evidence pipelines described
in Section~\ref{sec:Introduction} and the continuous post-market
surveillance settings where RWE plays an increasingly central role.

\paragraph*{Cross-cutting issues.}
Foundational challenges in data quality, interoperability, and bias
control cut across all methodological innovations. Integrating RCTs
with RWD/E requires harmonized data standards, transparent curation
pipelines, and rigorous evaluation of measurement error and missingness.
Biases, including confounding, selection bias, and temporal misalignment,
must be addressed explicitly, with causal diagrams and domain knowledge
guiding assumptions and model specification. Transparency, reproducibility,
and adherence to regulatory standards are not optional but necessary
for acceptance by agencies such as the FDA, EMA, NICE, and ICER. Open
science practices, including data sharing agreements, open-source
software, and reproducible workflows, should be embedded into the
research culture to ensure trust and credibility.

\subsection{What Statisticians Should Build and Learn}

\label{sec:future-build}

This section addresses \textbf{(Q4)} and \textbf{(Q5)}: how statisticians
should build trustworthy end-to-end evidence systems and how training
should evolve to support that role.

\paragraph*{For pharma and industry:}
Statisticians should build target-trial-ready RWD
pipelines with explicit design emulation; develop AI-assisted but
auditable phenotype and endpoint extraction workflows; create prespecified
transportability and external-control analysis plans; build validation
frameworks for digital twins and synthetic comparators; develop monitoring
systems for dataset shift, coding drift, and model drift; and create
evidence-integration platforms that combine prediction, inference,
and uncertainty quantification. A full-stack statistician
(Figure~\ref{fig:roadmap} Left) in industry today is expected to
lead target trial specification from the ground up, assess data provenance
and fitness for regulatory purposes, apply and develop doubly robust
causal inference and ML methods, oversee AI-assisted workflows
from phenotyping through validation, and communicate fluently with
regulatory agencies on integrative submissions, a profile that differs
substantially from the traditional clinical biostatistician. Evidence-platform
engineering is emerging as a unique statistical specialty, and pre-competitive
consortia for synthetic-control validation or digital-twin benchmarking
represent opportunities to shape methodological standards rather than
merely follow them.

\paragraph*{For regulatory science:}
Statisticians should define fit-for-purpose criteria
for RWD and validation standards for AI-derived variables and models;
clarify when dynamic borrowing is acceptable; require sensitivity
analysis and transparent estimands; define acceptable use cases and
evidentiary thresholds for synthetic data in regulatory submissions;
require that synthetic control arms undergo the same fitness-for-purpose
assessment as external RWD controls; and develop guidance on \textquotedbl good
synthetic data generation practices\textquotedbl{} analogous to existing
frameworks for AI/ML in medical devices \citep{FDA2025a}. The deeper
challenge, however, is structural. Many regulatory systems were designed
for a different era: one of discrete submissions, limited data sources,
and slow methodological change. AI and RWD now generate evidence continuously
and at a complexity that static frameworks struggle to accommodate.
What is needed is not a relaxation of standards but a redesign of
how they are structured: adaptive guidance that evolves with methodological
practice, pre-competitive qualification pathways for novel methods
such as synthetic controls and federated analyses, and tiered evidentiary
standards that match required evidence strength to decision stakes. 

\paragraph*{For training:}
The next generation of statisticians needs fluency
not only in causal inference and uncertainty quantification, but also
in the domains where they will actually apply these methods. A statistician
working with electronic health records, claims data, or clinical trial
systems who lacks grounding in clinical medicine, regulatory science,
or drug development will be technically capable but contextually lost:
unable to ask the right questions, detect implausible results, or
earn the trust of domain collaborators. Building this fluency demands
immersive, cross-sector training that statistics programs cannot provide
alone. Academia, industry, and government must collaborate to create
structured pathways, such as embedded internships, joint degree programs,
regulatory fellowships, and industry-academic consortia, that give
students genuine exposure to the environments where evidence is generated
and used. Students should encounter real data pipelines, real regulatory
constraints, and real decision pressures alongside their methodological
training, as core components of their education. On the technical
side, graduate curricula should be redesigned to include data engineering
and curation, multimodal data analysis, cloud and workflow tools,
AI/ML literacy beyond black-box use, and human-in-the-loop validation
of automated systems, alongside traditional statistical theory. The
2024 JSM town hall articulated this imperative under the banner of
the ``three Cs,'' namely communication, collaboration, and computation,
calling for curricula that incorporate deep learning, transformers,
graph-based methods, and intermediate software engineering \citep{donoho2026rebuilding}.
Given the central role of causal reasoning argued throughout this
paper, we propose a fourth C: causality. Without it, statisticians
risk conflating prediction with inference and lacking the vocabulary
to challenge AI-assisted analyses that overreach. The four Cs together
define the full-stack statistician's intellectual profile.


\subsection{A Stronger Statistical Identity in the Age of AI}

\label{sec:future-identity}

The broader implication is that statistics should
not define its role narrowly as a provider of inferential add-ons
to RWD or AI workflows. Instead, the field should assert a broader
identity as the discipline that governs credible learning from complex
data under uncertainty. That role includes causal reasoning, bias
assessment, uncertainty quantification, principled evidence integration,
and decision-oriented interpretation, but it now also requires deeper
engagement with data work, multimodal pipelines, scalable computation,
and real-world implementation. The field's future comparative advantage
lies not in competing with AI on predictive scale, but in leading
the construction of evidence systems that are interpretable, auditable,
robust, and suitable for real decisions.

In sum, the next phase of evidence generation depends on integration
rather than isolation. By combining the inferential strengths of statistics
with the predictive adaptability of AI/ML, and by bridging RCTs with
RWD/E under principled causal frameworks, the field can transform
current challenges into enduring opportunities. The ultimate goal
is to deliver evidence that is not only methodologically rigorous
and transparent but also clinically meaningful and directly relevant
to patient outcomes.

\section*{Acknowledgments}

We used ChatGPT to review grammar and refine the writing, and
NotebookLM to assist in the design and visual layout of Figure~1.
\if1\anon S.Y.'s research was partially supported by the
U.S. National Institutes of Health (grant \#1R01ES03627), the U.S.
Food and Drug Administration (grant \#U01FD007934), and the U.S. National
Science Foundation (grant \#SES 2242776). \fi

\if1\anon 
\section*{Disclosure statement}
M.G.\ is an employee of Pfizer and H.F.\ is an employee of Amgen. S.Y.\ reports there are no competing interests to declare.
 \fi









\bibliographystyle{agsm}
\bibliography{references}

\end{document}